\documentclass[amsmath,amssymb,aps,prl,reprint,superscriptaddress]{revtex4-1}

\usepackage{times}

\usepackage[squaren]{SIunits}
\usepackage{graphicx}
\usepackage{color}     

\usepackage{subfigure}
\def\ket#1{\mathinner{|{#1}\rangle}}

\newcommand{\sx}[0]{\ensuremath{\mathbf{\sigma}_x}}
\newcommand{\sy}[0]{\ensuremath{\mathbf{\sigma}_y}}

\begin{document}

\author{Yuting Ping}
\affiliation{Department of Materials, University of Oxford, Oxford OX1 3PH, United Kingdom}

\author{Brendon W. Lovett}
\affiliation{SUPA, School of Engineering and Physical Sciences, Heriot-Watt University, Edinburgh EH14 4AS, United Kingdom}
\affiliation{Department of Materials, University of Oxford, Oxford OX1 3PH, United Kingdom}

\author{Simon C. Benjamin}
\affiliation{Department of Materials, University of Oxford, Oxford OX1 3PH, United Kingdom}
\affiliation{Centre for Quantum Technologies, National University of Singapore, 3 Science Drive 2, Singapore 117543}

\author{Erik M. Gauger}
\email{erik.gauger@nus.edu.sg}
\affiliation{Centre for Quantum Technologies, National University of Singapore, 3 Science Drive 2, Singapore 117543}
\affiliation{Department of Materials, University of Oxford, Oxford OX1 3PH, United Kingdom}

\title{Practicality of spin chain `wiring' in diamond quantum technologies}

\begin{abstract}
Coupled spin chains are promising candidates for `wiring up' qubits in solid-state quantum computing (QC). In particular, two nitrogen-vacancy centers in diamond can be connected by a chain of implanted nitrogen impurities; when driven by a suitable global fields the chain can potentially enable quantum state transfer at room temperature. 
However, our detailed analysis of error effects suggests that foreseeable systems may fall far short of the fidelities required for QC. Fortunately the chain can function in the more modest role as a mediator of noisy entanglement, enabling QC provided that we use subsequent purification. For instance, a chain of 5 spins with inter-spin distances of \unit{10}{\nano\meter} has finite entangling power as long as the $T_2$ time of the spins exceeds \unit{0.55}{\milli\second}. Moreover we show that re-purposing the chain this way can remove the restriction to nearest-neighbor interactions, so eliminating the need for complicated dynamical decoupling sequences.
\end{abstract}

\maketitle

Spin chains with nearest neighbor XY coupling mediate coherent interactions between distant spin qubits with fixed locations, and can thus serve as channels to transfer quantum information~\cite{bose03,christandl04,yung05}. An important application would be to interconnect distant sub-registers of parallel parts in a scalable, solid-state quantum computer~\cite{lloyd93}, e.g.~in a diamond-based architecture at room temperature~\cite{stoneham09}. With an observed room-temperature coherence time of \unit{1.8}{\milli\second}~\cite{balasubramanian09}, the electron spin of individual nitrogen-vacancy (NV$^-$) defects in diamond is a promising candidate for a qubit~\cite{jelezko04}: Initialisation, coherent manipulation and measurement with nanoscale resolution ($\sim \unit{150}{\nano\meter}$) have already been experimentally demonstrated using optical techniques under ambient conditions~\cite{maurer10}. In addition, the long-lived $^{15}$N nuclear spin ($I = 1/2$) associated with each NV$^-$ center can act as a local, coherent memory, accessible via the hyperfine coupling~\cite{fuchs11, dutt07}. A universal set of quantum operations between the nuclear memory spin and the processing electronic spin qubit within each NV$^-$ center is available with microwave and radio-frequency pulses~\cite{cappellaro09, yao12, childress06}.
Two NV$^-$ centers with only a small separation ($r \lesssim \unit{10}{\nano\meter}$) may be entangled through direct electron spin dipole-dipole coupling as long as a $T_2$ time on the order of milliseconds can be maintained~\cite{neumann10}. However, individual addressability of the NV$^-$ center qubits demands larger separations of several tens or hundreds of nanometers~\cite{maurer10}, and the direct interaction becomes too weak.

\begin{figure}[h]
\begin{center}
\includegraphics[width=\linewidth]{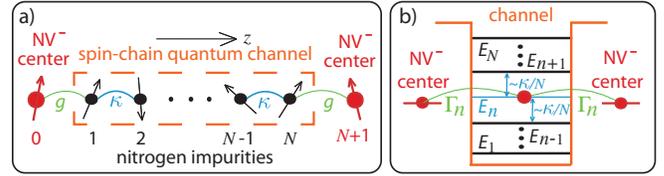}
\caption{(color online). (a) Two distant NV$^-$ centers are coupled through a spin chain consisting of $N$ nitrogen impurities; all spins are assumed to interact through nearest neighbor dipole-dipole couplings. The intra-chain and qubit-chain coupling strengths $\kappa$ and $g$, respectively, are determined by the geometrical arrangement of the defects, and the latter can be tuned through the NV$^-$ center ground state structure to enable high fidelity state transfer for $g \ll \kappa/\sqrt{N}$~\cite{yao11}.
(b) Illustration of the equivalent fermionic tunnelling picture: the channel possesses $N$ energy levels with spacings $|E_n - E_{n\pm1}| \sim \kappa/N$. When tuned into resonance with a channel level $E_n$, an NV$^-$ spin excitation tunnels through the channel and emerges on the opposite NV$^-$ center after a time $\tau_n = \pi / (\sqrt{2} \Gamma_n )$, where $\Gamma_n$ is the effective tunnelling rate. Off-resonant coupling to other levels is negligible provided that $\Gamma_n \ll |E_n - E_{n\pm1}|$ which is equivalent to $g \ll \kappa / \sqrt{N}$ (see text).}
\label{fig:system}
\end{center}
\vspace{-4mm}
\end{figure}

A recent proposal~\cite{yao11} suggested a chain of $N$ implanted nitrogen impurities (each with a ``dark'' electronic spin-1/2) as a coherent quantum channel to transfer quantum states between distant NV$^-$ centers at room-temperature (see Fig.~\ref{fig:system}a). Here, the electron spins of the NV$^-$ centers and the nitrogen impurities interact with each other through {\it nearest-neighbor} dipole-dipole coupling~\cite{epstein05, hanson06}. Importantly, the scheme does not require individual control of the chain spins, instead relying on global resonant driving fields to turn the effective Hamiltonian into an XY exchange model~\cite{yao12}. Reliable quantum state transfer ({\it QST}) between distant NV$^-$ centers can then be achieved through this spin-chain channel~\cite{yao11}. 

An important question is how well the above-described QST performs in the presence of realistic errors and imperfections~\cite{zwick11}. In the regime $g \ll \kappa/\sqrt{N}$, the QST has been found to be surprisingly robust against disorder in the intra-chain coupling $\kappa$~\cite{yao12b, zwick12}. The dominant error for the QST then arises due to the inevitable environmental decoherence of the spins in the channel. Focussing on such  quantum noise, we argue in this Letter that this is indeed the limiting factor, and that current diamond-based architectures will fail to meet quantum error-correction thresholds whilst still being able to support distributed quantum information processing.

We consider the following effective Hamiltonian for a chain consisting of $N$ spins plus two register spins located at either end of the chain (see Fig.~\ref{fig:system}): 
\begin{equation}
\hspace{-2.5mm} H_{\text{eff}} = \sum_{i=1}^{N-1} \kappa \sigma_+^i \sigma_-^{i+1} +\sum_{j=0, N} g \sigma_+^j \sigma_-^{j+1} + \mathrm{H.c.} ,
\label{eq:eff}
\end{equation}
where $\sigma^i_{\pm} = (\sx^i \pm i \sy^i)/2$ are the Pauli operators acting on spin $i$, the coupling strengths $\kappa$ and $g$ are proportional to the cubed inverse inter-spin-distance $1/r^3$, and H.c.~denotes the Hermitian conjugate. This Hamiltonian can be realised by applying global resonant microwave fields $H_{\text{drive}} = \sum_{i=0}^{N+1} \Omega_i\ \sx^i \cos \omega_i t $ with appropriate intensities $\Omega_i$ in the presence of a constant magnetic field applied in the $z$-direction. The $\omega_i$ denote the spin energy splittings including the  Zeeman and hyperfine components~\cite{yao12} (see Supplementary Information~\cite{sm} for more details). Importantly, the basis of the above Hamiltonian is {\it rotated} from the physical basis according to $(x, y ,z) \rightarrow (z, -y, x)$~\cite{yao12, sm}. Therefore, a $T_2$ ($T_1$) process acting on the physical spin corresponds to a spin-flip (phase-flip) error in the basis adopted for the Hamiltonian~(\ref{eq:eff}).

Following Ref.~\cite{yao11} we proceed by applying the  {\it Jordan-Wigner} transformation to Eq.~(\ref{eq:eff}), yielding
\begin{equation}
\label{eq:fermion}
\hspace{-2.5mm} H_{\text{eff}} = \sum_{i=1}^{N-1} \kappa\ c_i^{\dagger} c_{i+1}  + \sum_{j=0,N} g\ c_j^{\dagger} c_{j+1} + \mathrm{H.c.},
\end{equation}
with fermion creation and annihilation operators $c_i^{\dagger} := \sigma_+^i\ \text{exp} \big[- i \pi \sum_{j=0}^{i-1} \sigma_+^j \sigma_-^j\big]$, $c_i := \text{exp} \big[i \pi \sum_{j=0}^{i-1} \sigma_+^j \sigma_-^j\big]\ \sigma_-^i$, which observe the anticommutator relations $\{c_i, c_j^{\dagger}\} = \delta_{ij}$ and $\{c_i, c_j\} = 0 =\{c_i^{\dagger}, c_j^{\dagger}\}$~\cite{lieb61}. This transforms the spin degree of freedom into the presence or absence of a fermion at the relevant system site.

For $g \ll \kappa$ the coupling of the end NV$^-$ qubits to the channel $H'$ can be treated perturbatively~\cite{yao11}.  Diagonalising the first term of Eq.~(\ref{eq:fermion}) (and H.c.) yields
\begin{equation}
\tilde{H}_0 = \sum_{n=1}^{N} 2 \kappa \cos \frac{n\pi}{N+1}\ f_n^{\dagger} f_n, \\ \nonumber
\end{equation}
allowing us to write the second terms of Eq.~(\ref{eq:fermion}) as follows
\begin{equation}
\tilde{H}' = \sum_{n=1}^{N} \Gamma_n \left( c_0^{\dagger} f_n + (-1)^{n-1} c_{N+1}^{\dagger} f_n + \mathrm{H.c.} \right),
\label{eq:diagonal}
\end{equation}
where $f_n^{\dagger} = \sum_{j=1}^N \sin \frac{jn\pi}{N+1}\ c_j^{\dagger}\ /\sqrt{(N+1)/2}$ and $\Gamma_n = g \sin \frac{n\pi}{N+1} /\sqrt{(N+1)/2}$ for $n=1, 2, ..., N$~\cite{lieb61,wojcik05, venuti07}. The channel is now described by $\tilde{H}_0$ and possesses $N$ modes with energies $E_n = 2 \kappa \cos \frac{n\pi}{N+1}$. One can tune the energy of both NV$^-$ centers to $E_n$ by means of applying an appropriate detuning to $S^{0,N+1}_z$ in the rotated basis. This couples the NV$^-$ centers resonantly to the channel mode $n$ with a tunnelling rate $\Gamma_n$ (see Fig.~\ref{fig:system}b). Based on pure Hamiltonian evolution, the first full swap of the two end fermions occurs after a time
\begin{equation}
t_n = \frac{\pi}{\sqrt{2} \Gamma_n} = \frac{\pi \sqrt{N+1}}{2g\sin\frac{n\pi}{N+1}}.
\label{eq:time}
\end{equation}
The final `swapped' state acquires a controlled phase depending on the total number of fermions in the system~\cite{yao11, clark05}. Since for  $g \ll \kappa /\sqrt{N}$ off-resonant coupling to other channel modes is negligible, the phase arises from the single mode and, while generally unknown at finite temperature, is well-defined; it can then be corrected by employing a two-round protocol~\cite{yao12, yao11, markiewicz09, sm}. Based on this protocol, single-mode coupling QST is  {\it independent} of the initial chain state.

It is readily seen that the tunnelling rate $\Gamma_n$ reaches its maximal value for odd $N$ and $n=(N+1)/2$ with $E_n = 0$. In this case, no detuning is required, and the end NV$^-$ qubits resonantly couple to the zero-energy mode of the channel. We adopt this case of odd $N$ as the ideal implementation of the protocol. Since our calculations are not restricted to the single excitation subspace used in Ref.~\cite{yao12b}, we specifically consider the two cases of $N=3$ and $N=5$ resulting in simulations involving 6 and 8 spins (chain plus two registers and one ancilla) as explained further on. We simulate the full system dynamics by numerically integrating a Lindblad master equation~\cite{breuer02, sm} using Hamiltonian~(\ref{eq:eff}). Unless stated otherwise, we take the inter-spin spacing in the chain to be $r_{\text{N},\text{N}}=\unit{10}{\nano\meter}$ ($\kappa = \unit{26}{\kilo\hertz}$) while $g$ remains fully tuneable.

Numerous studies seek to determine a threshold below which scalable quantum computing is in principle possible. Typically, papers quote the threshold obtained by equating the error rates in state preparation, measurement, and qubit-qubit operations (generally, the latter are the most crucial). For qubits embedded in diamond and linked by spin chains, the most relevant thresholds are those for architectures where qubits are arranged on a lattice-like structure with each qubit being `wired' to only a few others. For this case, the threshold error rate is of order $1\%$: e.g. $0.75\%$ for the widely studied topologically protected cluster-state approach~\cite{raussendorf07a}, while $1.4\%$ can be obtained in {\it certain} circumstances~\cite{wang11}. Here we will take a target error rate of  $1\%$, i.e.~a fidelity requirement of $99\%$. Note that to avoid diverging resource requirements, one would not wish to build a computer with performance {\em near} the threshold; practically one might target an error rate ten times below the threshold~\cite{fowler12}. 
We demonstrate in the Supplementary Information~\cite{sm} that the threshold $1\%$ rate requires a physical $T_2$ time of \unit{16}{\milli\second} for the $N = 5$ chain when using an optimally tuned ratio of $g/\kappa$ \footnote{Interestingly, the required $T_1$ time, causing phase flips in the computational basis, may be much shorter, see Ref.~\cite{yao12b}}. By contrast, the longest measured $T_2$ time of the NV$^-$ center is an order of magnitude below this number at \unit{1.8}{\milli\second}~\cite{balasubramanian09}. Reported coherence times of the nitrogen impurity are much shorter (e.g.~\unit{5.5}{\micro\second} at room-temperature and~\unit{80}{\micro\second} at 2.5 K~\cite{takahashi08}). While this should be improvable, the nitrogen impurity is unlikely to substantially surpass the NV$^-$ center \cite{sm}. In reality, chains longer than $N=5$ will be desirable to properly separate the NV$^-$ centers. We therefore conclude that using QST as a fault-tolerant two-qubit operation may well be an infeasible target for any foreseeable technology.

Nonetheless, the suggested spin-chain quantum bus may still be able to support the distribution of entanglement, opening the possibility of employing distillation protocols to create high-fidelity entanglement over several runs~\cite{bennett96}. Recent studies show that highly imperfect inter-site links can be tolerated given that one has three or four qubits at each local site~\cite{li12, fujii12}. The key enabling property of the channel is then simply that it should be `quantum'  in contrast  to `classical', i.e. a channel that is  capable of transmitting a finite amount of entanglement with each run. Identifying the transition from quantum to classical channel with respect to realistic decoherence processes is the main purpose of this Letter. To address this question we attempt to transfer one half of a {\it Bell state} through the channel. More specifically, the {\it near} ($i=0$) NV$^-$ center starts off in a maximally entangled singlet state with an additional ancilla spin, $\ket{\Psi^-} = (\ket{0}_a\ket{1}_0 - \ket{1}_a\ket{0}_0)/\sqrt{2}$, and our observable is the entanglement of formation $E_F$~\cite{wootters98} between this ancilla and the {\it remote} NV$^-$ center at position $N+1$ \footnote{The phase acquired during the swap maps the singlet to a Bell equivalent state without degrading the quality of entanglement.}.

\begin{figure}[h]
\vspace{-2mm}
\begin{center}
\includegraphics[width=\linewidth]{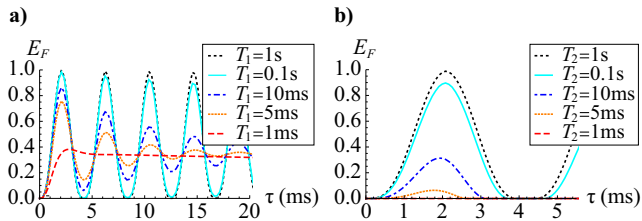}
\caption{(color online).  Entanglement of formation $E_F$ between the ancilla and the remote NV$^-$ center qubit as a function of the transfer time $\tau$ under an independent (a) spin-flip and (b) phase-flip  error model for the nitrogen spins ($N=3$). The errors are applied as Lindblad operators on each chain spin: (a)  $L_i = \sigma_{z_i}$ with rate  $\gamma = 1/T_1$; (b)  $L_i = \sigma_{x_i}$ with rate $\gamma = 1/T_2$. The initial state state is $\ket{\Psi^-}_{a0}\ket{000}\ket{0}_{N+1}$, and $g = \kappa / (10\sqrt{N})$.}
\label{fig:3sigmaxz}
\end{center}

\end{figure}

Let us start with $N=3$ as the shortest nontrivial odd chain \footnote{All channels with $N > 1$ possess more than one mode~\cite{christandl04}, so perfect QST for all values of $g$ is limited to $N=1$.}.  Fig.~\ref{fig:3sigmaxz}a shows that this channel is robust against physical spin-flip (i.e.~$T_1$-type~\footnote{We here take the inverse spin flip error rate as the $T_1$ time; this is not strictly the same as a $T_1$ time from energy relaxation.}) errors, which act as phase-flip errors in the basis of Eq.~(\ref{eq:eff}). A high degree of transferred entanglement is achievable for $T_1 > \unit{1}{\milli\second} $, and a finite amount of entanglement  survives in the presence of much larger error rates. Longer chains also remain robust against this type of error; for the single excitation subspace, we have simulated odd chains up to  $N = 21$  obtaining similar qualitative results~\cite{sm}.  Further, we have confirmed that this behavior is independent of the initial chain state, and that our results are in agreement with Ref.~\cite{yao12b}. 

On the other hand, dephasing of the physical spins, characterised by their $T_2$ time, induces spin-flip errors in the computational basis, with a much more damaging effect on the entangling power of the channel [see Fig.~\ref{fig:3sigmaxz}b]. For weak coupling between the NV$^-$ centers and the channel, the QST becomes classical, i.e.~$E_F$ vanishes completely, for $T_2 \leq \unit{3.2}{\milli\second}$. Applying the same model to the $N=5$ chain, the transition between a classical and a quantum channel occurs at $T_2 = \unit{7.6}{\milli\second} $. This {\it threshold} increases further for larger $N$ due to the transfer time becoming longer [see Eq.~(\ref{eq:time})],  as effectively more spins are exposed to the noise for longer.

\begin{figure}[ht]
\begin{center}
\includegraphics[width=\linewidth]{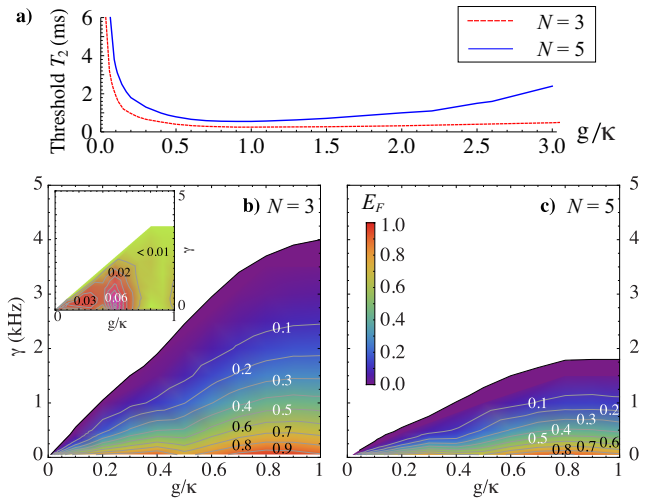}
\caption{(color online). (a) $T_2$ threshold separating classical   ($E_F = 0$) and quantum  ($E_F > 0$)  state transfer. The initial chain states are $\ket{\Psi^-}_{a0}\ket{000}\ket{0}_{N+1}$ and $\ket{\Psi^-}_{a0}\ket{00000}\ket{0}_{N+1}$, respectively, but the threshold for other initial states is similar. (b) and (c) Maximally achievable entanglement of formation $E_F$ between the ancilla spin and the remote NV$^-$ center spin as a function of $g/\kappa$ and  independent channel spin flip rate $\gamma = 1/T_2$ for $N=3$ and $N=5$. The channel is entirely classical in the white regions. The inset in (b) shows the variation in $E_F$ occurring for the eight different computational basis states of the channel spins in the $N=3$ case; in all cases the range of values of $E_F$ is small and always less than 0.07. Random checks for the $N=5$ case suggest a similar behavior for longer chains. }
\label{fig:3d3}
\label{fig:threshold}
\end{center}
\end{figure}

Unlike the high fidelity transfer of a particular quantum state, entanglement distribution does not benefit from coupling to only a single channel mode, allowing us to explore stronger coupling by varying $g$.  Fig.~\ref{fig:threshold}a shows the $T_2$ threshold which separates the quantum from the classical data bus for chains with $N=3$ and $N=5$, and we see that the highest tolerable dephasing rate occurs in the interval $g \in (0.8\kappa, \kappa)$. Expressed as a $T_2$ time threshold this gives \unit{0.25}{\milli\second} (\unit{0.55}{\milli\second}) for the $N=3$ ($N=5$) chain with  $r_{\text{N},\text{N}}=\unit{10}{\nano\meter}$. 

In Fig.~\ref{fig:3d3}b and~\ref{fig:3d3}c, we show the entangling capacity of the $N=3$ and $N=5$ channels as a function of  $g$ and the dephasing rate $\gamma = 1/ T_2$. We note that high channel entangling power is only realised for a small decoherence rate of the nitrogen spins, even for larger $g$. Unsurprisingly, the channel is highly inefficient close to the classical threshold. 

To `connect' two NV$^-$ centers separated by a fixed distance of \unit{40}{\nano\meter}, we consider a chain comprising three and five nitrogen spins. In both cases the register spins are assumed to be \unit{10}{\nano\meter} away from the ends of the chain, so that the chain's inter-spin distances are $r_{\text{N},\text{N}} = \unit{10}{\nano\meter}$ and $r_{\text{N},\text{N}} = \unit{5}{\nano\meter}$, respectively. The first case then corresponds to the $N=3$ chain we have studied so far, but for the $N=5$ case we obtain a minimal $T_2 \approx \unit{70}{\micro\second}$ for quantum communication (for $g = \kappa$) which is significantly shorter than \unit{0.25}{\milli\second} for the $N=3$ chain. The benefit of stronger coupling therefore outweighs the drawback of a longer chain for a fixed separation between the register spins by speeding up the transfer process. However, how many impurity spins can be deployed will crucially depend on the achievable implantation precision, since even sub-nanometer imperfections entail significant coupling disorder for closely spaced chains.

So far, we have assumed nearest-neighbor couplings [see Eq.~(\ref{eq:eff})], which could be realised through dynamical decoupling~\cite{yao12}. In practice, one needs a coupling strength `high pass filter', which would typically reduce effective coupling strengths and add further complexity to experimental implementations. We now relax this assumption and include all pairwise couplings with a dipole-dipole interaction $1/r^3$ distance dependence (similar to Refs.~\cite{kay06, ronke11}), including register to chain spin couplings. 
\begin{figure}[h]
\begin{center}
\includegraphics[width=0.99\linewidth]{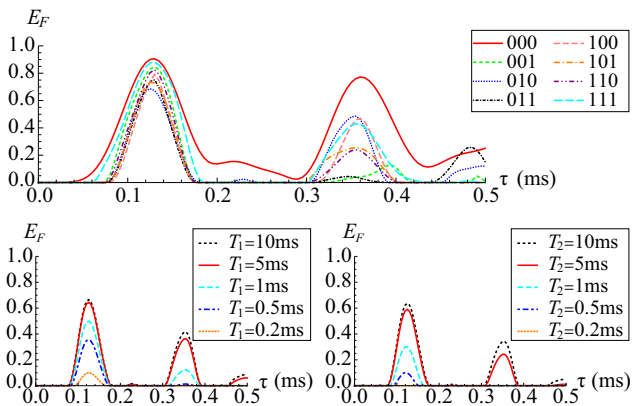}
\caption{(color online). Top: $E_F$ between the ancilla spin and the remote NV$^-$ center as a function of the transfer time $\tau$ for all eight initial chain states of the $N=3$ non-nearest neighbor chain. All spins are assumed equally spaced, with the coupling strength for nearest neighbors being \unit{26}{\kilo\hertz}. Bottom: $E_F$ for the initial state  $\ket{\Psi^-}_{a0}\ket{010}\ket{0}_{N+1}$ with decoherence processes of independent phase-flip noise (left) and  spin-flip noise (right) in the basis of Hamiltonian~(\ref{eq:eff}) corresponding to physical $T_1$ and $T_2$ processes, respectively.
}
\label{fig:nnn}
\label{fig:3nnnsigmaxz}
\end{center}
\end{figure}

Fig.~\ref{fig:nnn} demonstrates that the entanglement transfer for a uniformly spaced spin chain now depends on the initial chain state. However, the first  maximum of all initial states coincides and gives rise to a reasonable degree of transferred entanglement. Thus we pick the initial chain state  $\ket{\Psi^-}_{a0}\ket{010}\ket{0}_{N+1}$ as an example that is indicative of the performance to be expected. Interestingly, $T_1$ and $T_2$ type processes degrade the $E_F$ with much more similar severity compared to the nearest neighbor only coupling case \footnote{Restoring nearest neighbor coupling only for the register spins largely reintroduces the asymmetric behaviour with respect to the two different types of noise.}. Nonetheless, there still exists a threshold in the $T_2$ coherence time, and to ensure quantum communication for an arbitrary initial state, we require $T_2 > \unit{0.28}{\milli\second}$, which is only slightly longer than for the nearest neighbor chain. For the case of the $N=5$ chain (not shown), we have determined the transition threshold to be $T_2 > \unit{0.67}{\milli\second}$, which again only represents a modest increase. Taking into account the complexity and expected reduction in effective coupling strengths from dynamical decoupling sequences, it may thus be more practical to drop the restriction of only nearest neighbor interactions. As we show in the Supplementary Information, the state transfer still remains similarly robust to small coupling-strength disorder~\cite{sm}.

In conclusion, by employing numerical simulations we have studied the impact of inevitable decoherence processes on the entanglement capacity of a spin-chain bus that is realised by dipole-dipole coupling of crystal defects. Limiting our discussion to chains of length $N=3$ and $N=5$ as likely candidates for a first experimental demonstration (and also for numerical tractability) has allowed us to obtain insight into important characteristics of such a protocol in the presence of realistic noise. Our conclusions will be equally relevant for longer chains, which are known to be even more susceptible to decoherence~\cite{cai06, hu09, zeng09}.
We have shown that directly meeting quantum error correction thresholds remains infeasible even if all spins possessed the exceptional coherence time of NV$^-$ centers. In contrast, the distribution of a finite amount of entanglement appears realistic with current systems, offering the possibility of applying distillation protocols to boost the transmitted entanglement with additional local operations and classical communication~\cite{bennett96}. 

\begin{acknowledgements}
The authors are grateful to Earl Campbell and Jason Smith for useful discussions. This work was supported by the National Research Foundation, the Ministry of Education, Singapore. BWL acknowledges the Royal Society for a University Research Fellowship. YP thanks Hertford College, Oxford for a scholarship.

\end{acknowledgements}

\end{document}


\author{Yuting Ping}
\affiliation{Department of Materials, University of Oxford, Oxford OX1 3PH, United Kingdom}

\author{Brendon W. Lovett}
\affiliation{SUPA, School of Engineering and Physical Sciences, Heriot-Watt University, Edinburgh EH14 4AS, United Kingdom}
\affiliation{Department of Materials, University of Oxford, Oxford OX1 3PH, United Kingdom}

\author{Simon C. Benjamin}
\affiliation{Department of Materials, University of Oxford, Oxford OX1 3PH, United Kingdom}
\affiliation{Centre for Quantum Technologies, National University of Singapore, 3 Science Drive 2, Singapore 117543}

\author{Erik M. Gauger}
\affiliation{Centre for Quantum Technologies, National University of Singapore, 3 Science Drive 2, Singapore 117543}
\affiliation{Department of Materials, University of Oxford, Oxford OX1 3PH, United Kingdom}

\title{Supporting Information for Practicality of spin chain `wiring' in diamond quantum technologies}

\begin{abstract}
In this Supporting Information document, we present further information and calculations supporting the conclusions of the main Letter. In particular, we discuss the level structure of the NV$^{-}$ center, and give an explicit derivation of the XY spin chain Hamiltonian used in the main text. Furthermore, we show that fault-tolerant two-qubit gates require unrealistically long $T_2$ times. Finally, we present our results for $T_1$ processes in longer chains and the effects of coupling strength disorder.
\end{abstract}

\maketitle

\renewcommand{\thefigure}{S\arabic{figure}}
\renewcommand{\theequation}{S\arabic{equation}}

\section{Electronic spin qubit of NV$^-$ defects}
\begin{figure}[h]
\vspace{2mm}
\begin{center}
\includegraphics[width=2.3in]{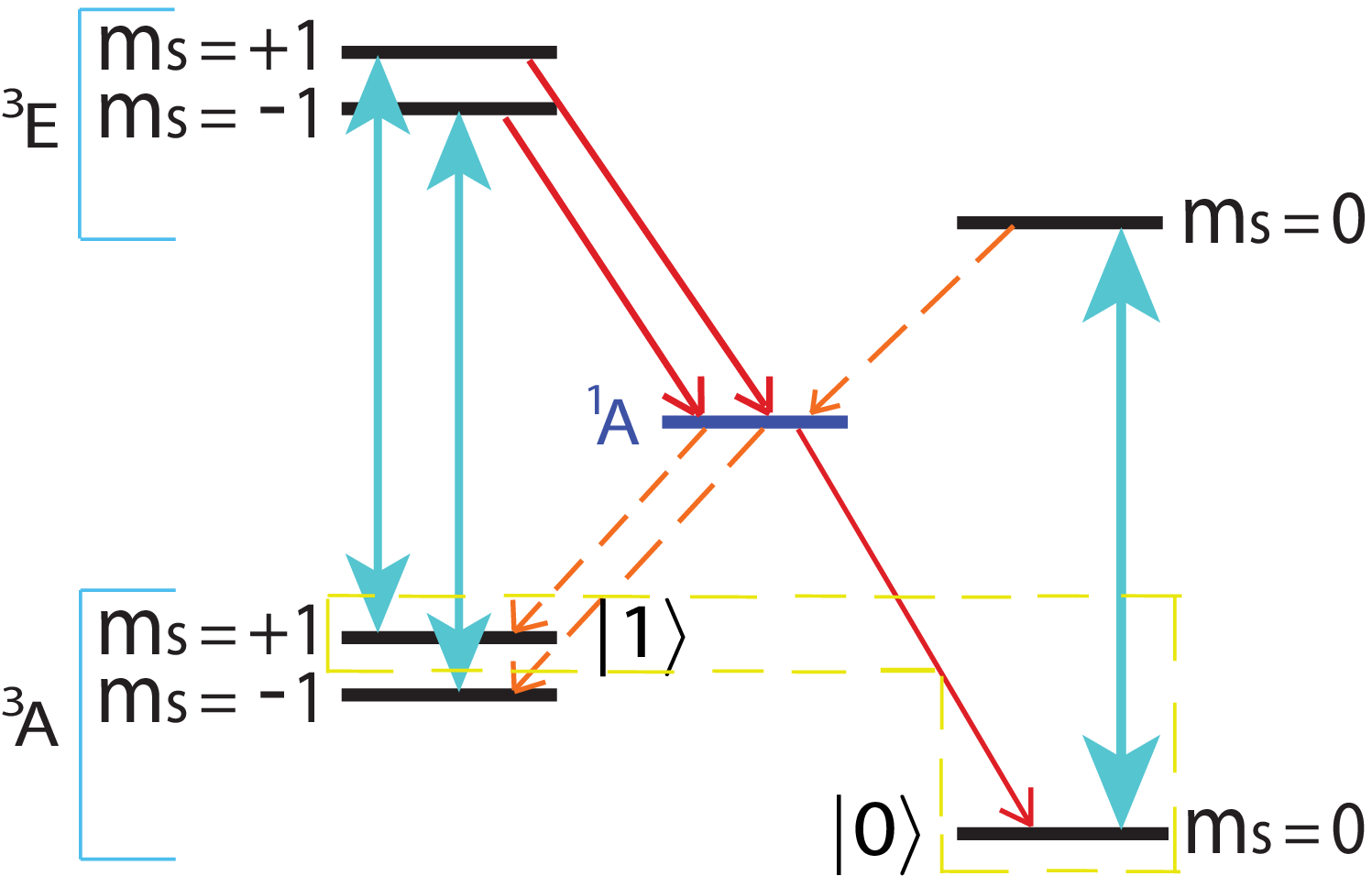}
\caption{Simplified electronic structure of the NV$^{-}$ colour center. The zero-field splitting between the $m_s = 0$ and the $m_s = \pm 1$ ground triplet is \unit{2.88}{\giga \hertz}, and transitions between different spin levels can be effected through resonant microwave pulses~\cite{oort88, redman91, jelezko04}. The $^{1}$A singlet level is metastable and provides a route for intersystem crossing relaxation (red arrows) from the excited triplet to the ground state: Thick red lines have associated ISC rates that are several orders of magnitude faster than the thinner red line, and dashed transitions can be considered to be negligible~\cite{manson06, rogers08}. }
\label{fig:levels}
\end{center}
\end{figure}

The nitrogen-vacancy colour defect in diamond consists of a substitutional nitrogen atom and an adjacent vacancy. In its negatively charged state, the NV$^-$ center traps an excess electron and possesses a paramagnetic ground state ($S=1$) with extraordinarily long spin lifetime. For each NV$^-$ center, the spin-triplet ground state $^3$A consists of the $m_s = 0$ and the degenerate (by C$_{3v}$ symmetry) $m_s = \pm 1$ sublevels, split by \unit{2.88}{\giga \hertz}~\cite{oort88, redman91}. An external magnetic field lifts the degeneracy between the $m_s = + 1$ and the $m_s = -1$ sublevels. The \unit{637}{\nano\meter} optical transitions between the ground and the excited $^{3}$E triplet states are predominantly spin-conserving~\cite{davies76, tamarat08}, but an intersystem crossing (ISC) occurs via the singlet $^1$A state~\cite{manson06, rogers08} (see Fig.~\ref{fig:levels}). 

\vspace{-2mm}
To use the electron spin of an NV$^-$ center as a  quantum bit, one can, for example, encode the states $\ket{0}$ and $\ket{1}$ into the $m_s = 0$ and $m_s = +1$ sublevels of the ground triplet, respectively. Optical pumping then polarizes the qubit into the $\ket{0}$ state, and resonant microwave pulses enable single qubit operations, such as the creation of a coherent superposition state $(\ket{0} +  \ket{1}) / \sqrt{2}$~\cite{jelezko04, maurer10}. The spin can also be read out optically with a fluorescence technique~\cite{jelezko04, maurer10}.

\section{Effective System Hamiltonian} 
Following Ref.~\cite{yao12} we derive the effective system Hamiltonian for the chain shown in Fig.~1 of the main text. We assume the presence of a constant magnetic field of strength $B$ in the $z$-direction which is defined by the symmetry axis of the NV$^-$ center, i.e.~the [111] crystal axis. The Hamiltonians for individual NV$^-$ and N defects are given by~\cite{childress06, hanson06}, respectively,
\begin{align}
H_{\text{NV}}	& =  g_e \mu_B B S^{\text{NV}}_z - g_n \mu_n B I_z  \nonumber \\
	         & +  D (S^{\text{NV}}_z)^2 + A_{\text{NV}} {\mathbf I} \cdot {\mathbf S^{\text{NV}}}, \label{eq:NVraw} \\ 
H_{\text{N}}          & =  g_e \mu_B B S_z - g_n \mu_n B I_z + {\mathbf S} \cdot{\mathbf A}_{\text{N}} \cdot {\mathbf I}~, \label{eq:Nraw}
\end{align}
where $g_e$ ($g_n$) is the electron (nuclear) g-factor, $\mu_B$ ($\mu_n$) is the Bohr (nuclear) magneton, $A_{\text{NV}}$ (${\mathbf A}_{\text{N}}$) denotes the hyperfine coupling constant (tensor),  $D$ is the zero-field splitting for the NV$^-$ center, and ${\mathbf S}^{(\text{NV})}$ (${\mathbf I}$) is the full electronic (nuclear) spin operator.

For each NV$^-$ center, we encode the qubit basis states $\ket{0}$ and $\ket{1}$ in the $m_s = 0$ and $m_s = +1$ sublevels of the ground triplet (see Fig.~\ref{fig:levels}), respectively. Expressed in the computational basis of the qubit, we can approximate Eq.~(\ref{eq:NVraw}) as:
\begin{equation}
H^{\text{qubit}}_{\text{NV}} =  \frac{\omega^{\text{NV}}_0}{2} \sz + \frac{A_{\text{NV}}}{2} I_z \sz~,
\label{eq:HNV}
\end{equation}
where the electronic Zeeman energy $\omega^{\text{NV}}_0 := D + g_e \mu_B B $, and $\sz$ is the usual Pauli $z$-operator acting on the qubit. The effective Hamiltonian above does not include the heavily suppressed hyperfine spin-flip terms, nor the negligible nuclear Zeeman term ($A_{\text{NV}}, g_n \mu_n B \ll \omega^{\text{NV}}_0$)~\cite{childress06}. 

For the $i^{\text{th}}$ nitrogen defect with $S=1/2$, Eq.~(\ref{eq:Nraw}) can be similarly approximated as
\begin{equation}
H^{i}_{\text{N}} =  \frac{\omega_0}{2} \sz^i + \frac{A^i_{\text{N}\parallel}}{2} I^i_z \sz^i~,
\label{eq:HN}
\end{equation}
where $\omega_0 := g_e \mu_B B \approx \unit{10}{\giga\hertz}$, and $\sigma^i_z$ acts on the spin qubit of the $i^{\text{th}}$ N impurity. The hyperfine couplings $A^i_{\text{N}\parallel}$ depend on the {\it Jahn-Teller} orientation of each N defect, and can take two possible values $- 118.9$ MHz and $- 159.7$ MHz~\cite{cox94, kedkaew08, yao12}. 

The magnetic dipole-dipole interaction between two electron spins $i$ and $j$ (both N defects, or one N and one NV$^-$ center) is generically given by
\begin{align}
H^{ij}_{\text{dip}} &= \frac{\mu_0 g^2_e \mu^2_B}{4 \pi r^3} \left( {\mathbf S^i} \cdot {\mathbf S^j} - 3 ({\mathbf S^i} \cdot \mathbf{\hat{r}})({\mathbf S^j} \cdot \mathbf{ \hat{r}}) \right) \nonumber \\
&= \frac{\mu_0 g^2_e \mu^2_B}{4 \pi r^3} \left(S_x^iS_x^j + S_y^iS_y^j - 2 S_z^iS_z^j\ \right) \nonumber \\
& \simeq - \frac{\mu_0 g^2_e \mu^2_B}{2 \pi r^3} S_z^iS_z^j~,
\label{eq:dipole}
\end{align}
where $\mu_0$ is the vacuum magnetic permeability, $r$ is the separation between the spins, and $\mathbf{\hat{r}}$ denotes the unit vector connecting the two spins, here assumed to be parallel to the $z$-direction (as is the case in Fig.~1 of the main text). In the last step, we have neglected the spin-flip terms as before~\footnote{In this study we consider a chain that is aligned with the external magnetic field (the $z$-direction). This leads to an effective dipole coupling strength that is increased by a factor of two compared to~Ref.~\cite{yao12}. However, to support register architectures with the desired two- or three-dimensional lattices ~\cite{yao12}, connections perpendicular to the applied field will also be required.}, since $\mu_0 g^2_e \mu^2_B / (4 \pi r^3) \simeq \unit{52}{\kilo\hertz} \ll A_{\text{NV}}, A^i_{\text{N}\parallel}$ for a spacing of $r= \unit{10}{\nano\meter}$~\cite{childress06, yao12}.  We can therefore write the Hamiltonian for a pair of spins as~\cite{cordes87} 
\begin{align}
H_{\text{N},\text{N}} & = \kappa\ \sigma_z^1 \sigma_z^2 + \sum_{i=1,2} \frac{( \omega_0 + \delta_i )}{2}\ \sigma_z^i , \label{eq:HNN} \\
H_{\text{N},\text{NV}} & = g\ \sigma_z^0 \sigma_z^1 + \frac{( \omega^{\text{NV}}_0 + \delta )}{2}\ \sigma_z^0 + \frac{(\omega_0 + \delta_1)}{2}\ \sigma_z^1~, \label{eq:HNNV}
\end{align}
where the hyperfine terms $\delta := \pm A_{\text{NV}}/2$ and $\delta_i := \pm A^i_{\text{N}\parallel}/2$ ($i= 1, 2, ..., N$); the dipolar coupling strengths are $\kappa := - \mu_0 g^2_e \mu^2_B/ (8 \pi r_{\text{N},\text{N}}^3)$ and $g := - \mu_0 g^2_e \mu^2_B/(8 \pi r_{\text{N},\text{NV}}^3)$. 

Eqns.~(\ref{eq:HNN}) and (\ref{eq:HNNV}) are easily generalised to a nearest-neighbor coupled chain. We assume that the entire chain is driven by the following resonant global fields
\begin{align}
H_{\text{drive}} & = \sum_{i=1}^{N} \Omega\ \sx^i \cos \omega_i t \nonumber \\
& + \sum_{j=0, N+1} \Omega_0\ \sx^j \cos (\omega^{\text{NV}}_0 + \delta) t ~,
\label{eq:driving}
\end{align}
where all four possible frequencies $\omega_i = \omega_0 + \delta_i$ [see Eqns.~(\ref{eq:HN}),~(\ref{eq:HNN}) and~(\ref{eq:HNNV})] are applied to address the nitrogen impurities~\cite{yao12}. The field intensities $\Omega_0, \Omega$ are chosen to fit in the hierarchy $\vert \kappa \vert \ll \Omega, \Omega_0 \ll \omega_i, \omega^{\text{NV}}_0 + \delta$. 

Making a rotating wave approximation in the usual rotating frame (see Appendix) and adopting the rotated basis $(x, y ,z) \rightarrow (z, -y, x)$~\cite{yao12} then yields an effective XY interaction model for the chain:
\begin{align}
\hspace{-2.5mm} H_{\text{eff}}  & = \sum_{i=1}^{N-1} \kappa (\sigma_+^i \sigma_-^{i+1} + \sigma_-^i \sigma_+^{i+1}) \nonumber \\
& +\sum_{j=0, N} g (\sigma_+^j \sigma_-^{j+1} + \sigma_-^j \sigma_+^{j+1}) ~,
\label{eq:eff}
\end{align}
where $\sigma^i_{\pm} = (\sx^i \pm i \sy^i)/2$. 

Eq.~({\ref{eq:eff}}) is the effective system Hamiltonian given as Eq.~(1) of the main text. Controlling the magnitude $\Omega_0$ can also effectively tune the coupling $g$ between the NV$^-$ center spins and the N defect spin chain~\cite{yao12}. Note that $\kappa$ and $g$ in Eq.~(\ref{eq:eff}) are negative, however, we shall take absolute values for both couplings, since any global phase for the entire spin chain due to the sign of the interaction is irrelevant.

\section{Decoherence model}

We model the time evolution of our system with a standard Lindblad master equation~\cite{breuer02} :
\begin{align}
\dot{\rho} =	&-i \left[ H_{\text{eff}}, \rho \right] \nonumber \\
	 		&+ \sum_{i=1}^{N} \gamma_i \left(  L_i \rho L_i^{\dagger}   - \frac{1}{2} \left( L_i^{\dagger} L_i \rho + \rho  L_i^{\dagger} L_i \right) \right)~, 
			\label{eq:me}
\end{align}
where $\rho$ is the density matrix of the channel including the NV$^{-}$ center register spins, $H_{\text{eff}}$ is the effective system Hamiltonian~(\ref{eq:eff}), the $\gamma_i$ are the noise rates and the $L_i$ the noise operators. 

For spin-flip noise (i.e. $T_1$-like processes) all noise rates are $\gamma \equiv \gamma_i = 1 / T_1$ and we use noise operators $L_i = \sigma_z^i$ (since the computational basis is rotated from the physical basis according to $x \to z$). There is one operator for each N impurity acting independently on its spin, a choice corresponding to the case where the source of the noise is predominantly local to each spin. This reflects the spatial extent of the spin chain, in which each channel spin can be considered as interacting with its own environment, e.g.~the nuclear spin bath, nearby defect sites and local phonons due to lattice distortion. Similarly, for pure dephasing noise, we use $L_i = \sigma_x^i$ with associated rates $\gamma = 1 / T_2$.

Interestingly, Eq.~(\ref{eq:me}) only involves the three parameters $g, \kappa$ and $\gamma$; and the dynamics of the systems is invariant under a suitable rescaling of the coupling strength, unit of time and noise rate. Therefore,  coherence time thresholds can be easily obtained for coupling strengths different from the ones presented in this paper. For example, if one is interested in  $g$ and $\kappa$ that are only half as large, the noise rate simply needs to be halved and the corresponding coherence time doubled. This rescaling then gives rise to the same entangling capacity $E_F$ of the channel, and also correctly captures the quantum to classical transition.

The $T_2$ time for the N defect spins is unlikely to substantially exceed that of the NV$^{-}$ centers, since the coherence time of both types of spin is ultimately limited by the same physical processes: interaction with additional electronic defect spins and with the nuclear spin bath~\cite{benjamin09}. Experimental evidence for the NV$^{-}$ and N spin $T_2$ being limited by the same spin bath has been reported in Ref.~\cite{takahashi08}.

Optimal control strategies can mitigate the loss of coherence due to a small set of interacting quantum systems, e.g.~small environments of up to six additional spins were addressed by Ref.~\cite{grace07}. However, as the complexity and size of the environment increase, these techniques become more difficult to implement and the coherence time will necessarily begin to decrease \cite{grace07}. Optimal control can thus help to combat one source of decoherence but will not be able to overcome other unavoidable (Markovian) decoherence channels supported by the relatively large diamond crystal required for the envisaged architecture.

\section{$T_2$ thresholds for fault tolerance} 

We simulate the quantum state transfer (QST) for spin chains with odd $N$ in the weak coupling regime, $g=\kappa / (10\sqrt{N})$, and for a \unit{10}{\nano\meter} intra-chain spacing. After the transfer the acquired phase $\pm 1$ of the target state is corrected to enable a direct comparison with the initial state (in practice the phase would be cancelled by employing a two-round protocol~\cite{yao12, markiewicz09, yao11}). To evaluate the fidelity of the transfer process, we use the measure $F^2(\rho, \sigma) = \text{Tr} \left(\sqrt{\sqrt{\rho} \sigma \sqrt{\rho}} \right)^2$~\cite{jozsa94}. The N defect spins are subjected to independent physical $T_2$ processes, realised as spin flips in the computational basis.

\begin{figure}[h]
\vspace{3mm}
\begin{center}
\includegraphics[width=3.4in]{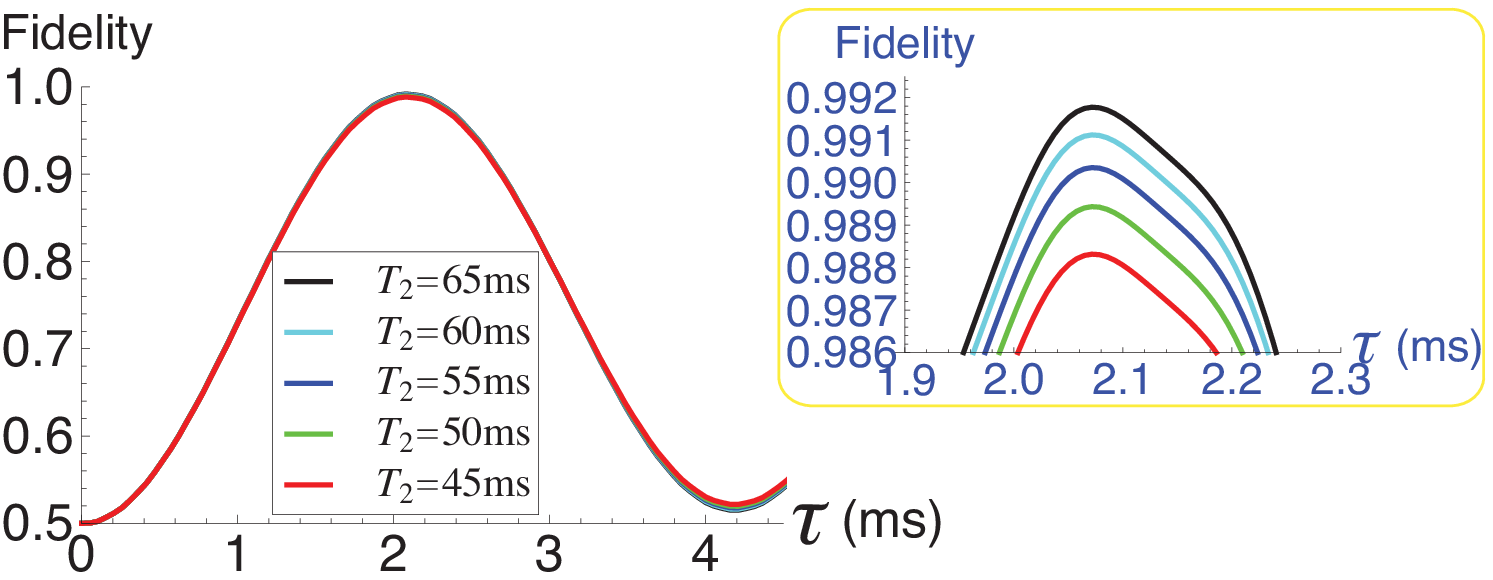}
\caption{Fidelity $F^2$ of the transferred state $\rho$ with respect to the input $\sigma = \ket{+}\bra{+}$, through the $N=3$ chain as a function of the transfer time $\tau$. The initial state is $\ket{+}_0\ket{000}\ket{0}_{N+1}$, and we operate in the weak coupling regime $g=\kappa/(10\sqrt{N})$ for $\kappa = \unit{26}{\kilo\hertz}$. The nitrogen spins experience independent dephasing at a rate $\gamma = 1/T_2$. Whilst the curves seem to coincide in the main plot, the inset shows a zoomed-in view near the maximum on the scale relevant for fault tolerance.}
\label{fig:ftqc}
\end{center}
\end{figure}

Fig.~\ref{fig:ftqc} shows that meeting a fault-tolerance threshold of order $F^2 \geq 99\%$ 
requires unrealistically long coherence times of the defect spins. More specifically, the shortest non-trivial $N=3$ chain achieves a sufficiently high fidelity only for $T_2 \geq \unit{54}{\milli\second}$ whereas this number increases to $T_2 \geq \unit{88}{\milli\second}$ for $N=5$.

As described in the main text, in the weak coupling regime NV$^-$ center excitations tunnel through the (single) zero-energy mode of the chain, and off-resonant coupling to other modes is negligible. This enables high-fidelity quantum state transfer for long enough coherence times. However, because transfer time $\tau \sim 1 / g$ is longer for weaker $g$, the state transfer is more susceptible to decoherence, and it may thus be advantageous overall to use stronger $g$ at the cost of coupling to several modes and sacrificing some theoretical fidelity. In the following table we list the $T_2$ times in milliseconds required for achieving an error rate below $1\%$ for different $g / \kappa$ ratios with $\kappa = \unit{26}{\kilo\hertz}$:
\begin{center} 
\begin{tabular}{ |c | |c|c |c|c |c|c |c|c |c|c |c| }
  \hline                        
  $g/\kappa$ & $0.1 / \sqrt{N}$ & 0.1 & 0.2 & 0.3 & 0.4 & 0.5 & 0.6 & 0.7 & 0.8 & 0.9 & 1 \\
  \hline 
   $N=3$ & 54 & 31 & 77 & -- & 19 & -- & -- & 31 & 6.8 & 10 & -- \\
  \hline  
   $N=5$ & 88 & 43 & 30 & 25 & -- & -- & -- & 28 & 16 & -- & -- \\
  \hline  
\end{tabular}
\end{center}

\noindent where  `--' denotes that the desired fidelity is never achieved for the initial states $\ket{+}_0\ket{000}\ket{0}_{N+1}$ and $\ket{+}_0\ket{00000}\ket{0}_{N+1}$, respectively. The fluctuating behavior seen in this table is consistent with the results reported in Ref.~\cite{yao11}. We conclude that for the studied chains fault-tolerant quantum computation demands coherence times of several milliseconds for the $N=3$ chain and a few tens of milliseconds for the more interesting $N=5$ chain, even when dropping the weak coupling constraint and in the absence of any other imperfections.

\section{$T_1$ Process in Longer Chains} 
In the main text, we show the effects of $T_1$ and $T_2$ processes for chains of length $N=3$ and $5$. Considering only $T_1$ processes and restricting ourselves to the zero and single excitation (computational) subspace \footnote {i.e.~no more than one spin may be found to be in the $\ket{1}$ while all others are in the $\ket{0}$ state if a measurement were performed} significantly reduces the numerical complexity, allowing us study longer chains. 

\begin{figure}[h]
\begin{center}$
\begin{array}{cc}
\hspace{-1.1mm} \subfigure[\ $g=\kappa$]{\includegraphics[width=0.495\linewidth]{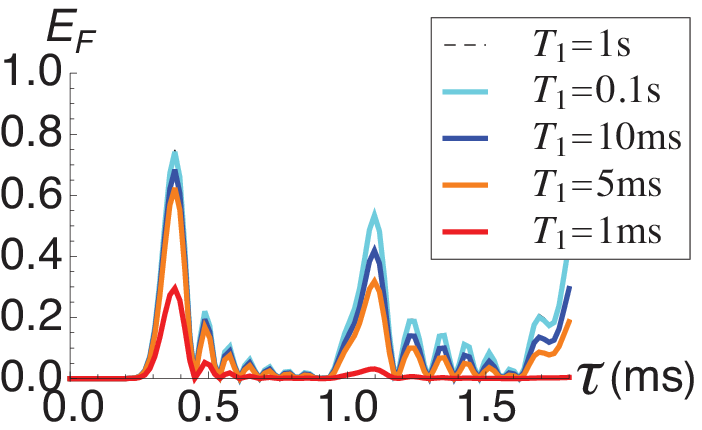}} &
\hspace{-0.9mm} \subfigure[\ $g=\frac{\kappa}{10\sqrt{N}}$]{\includegraphics[width=0.495\linewidth]{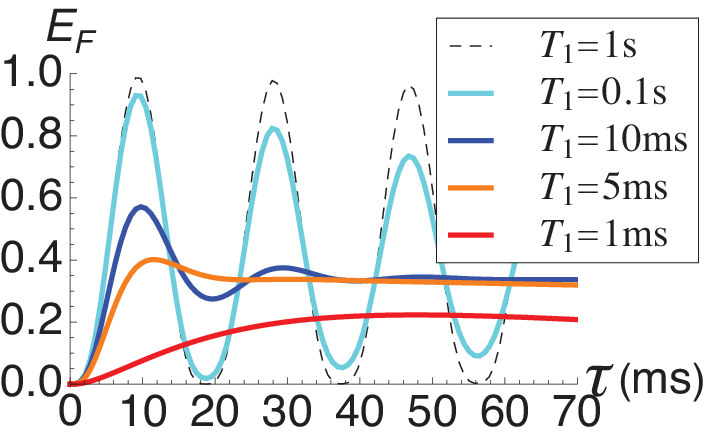}}  
\end{array}$  
\caption{Entanglement of formation $E_F$ between the ancilla spin and the remote NV$^-$ qubit as a function of the transfer time $\tau$ under an independent, physical bit-flip model for nitrogen spins with rate $\gamma = 1/T_1$. The chain length is $N=15$, and all chain spins are initially in the state $\ket{0}$. Only nearest-neighbor interactions are included.}
\label{fig:long}
\end{center}
\end{figure}

Fig.~\ref{fig:long} illustrates that a sizeable amount entanglement can be transmitted through a $N=15$ chain for $T_1$ times as short as a millisecond, both for strong and weak coupling of the NV$^-$ centers to the chain. When the bit flip rate gets small, $T_1 \gtrsim \unit{0.1}{\second}$, the weak coupling approach possesses a much higher entangling power, despite taking substantially longer. For $T_1 \sim \unit{10}{\milli\second}$~\cite{maurer12}, however, both approaches are comparable. Taking into account the increased robustness of the strong coupling case against $T_2$ processes (see the main text) suggests that overall $g \approx \kappa$ is likely the better choice for tackling decoherence.

\begin{figure}[h]
\vspace{-4mm}
\begin{center}
\includegraphics[width=2.6in]{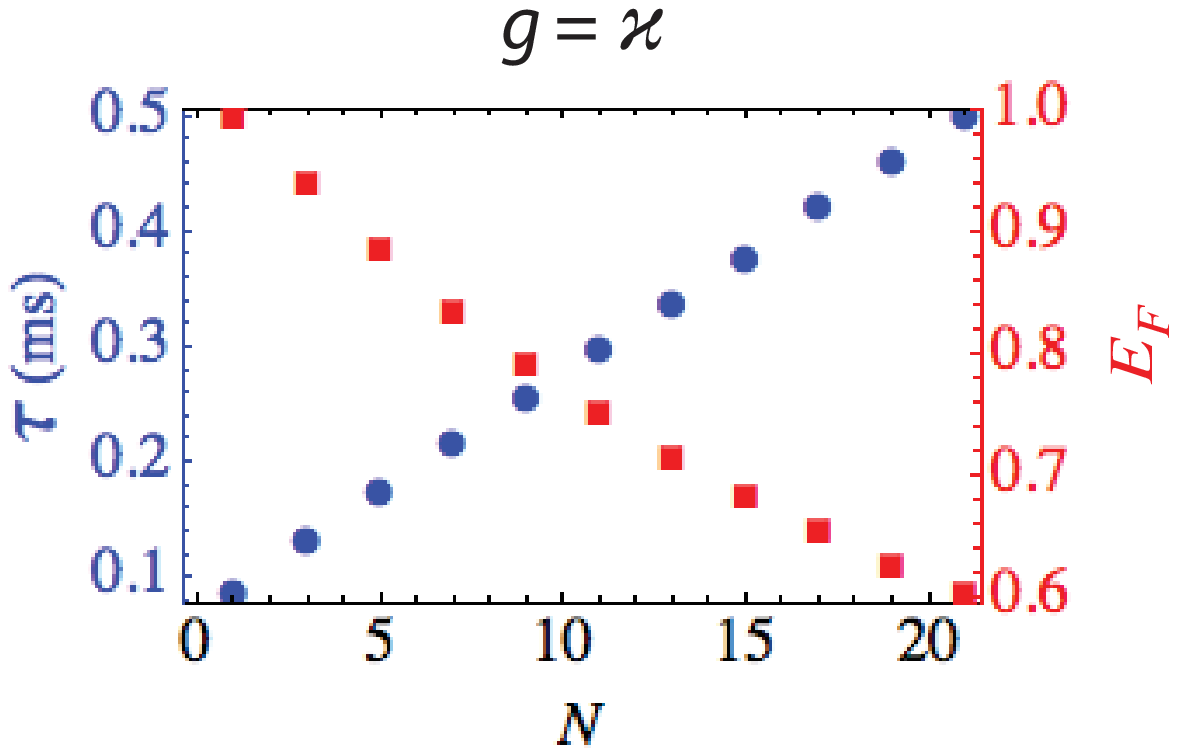}
\vspace{-3mm}
\caption{Plot of the transfer time $\tau$ (the first peaks in plots like Fig.~\ref{fig:long}(a)) and the resulted $E_F$ values between the distant spin and the $(N+1)^{\text{th}}$ NV$^-$ qubit as a function of the channel length $N$, with fixed intra-chain spacing $r_{\text{N,N}} = 10$nm ($g=\kappa = \unit{26}{\kilo\hertz}$), under an independent phase-flip model (in the rotated basis $x \leftrightarrow z$) on each nitrogen spin with practically relevant rate $\gamma = 1/T_1 = \unit{100}{\hertz}$ ($T_1= \unit{10}{\milli\second}$). The channel spins are initially all in the state $\ket{0}$, and interact with their nearest neighbors only.} 
\label{fig:trend}
\end{center}
\end{figure}

Fig.~\ref{fig:trend} considers different chain lengths with $g=\kappa$ coupling, showing the optimal transfer duration $\tau$ and maximally achievable $E_F$ for each case under an independent bit flip model with $T_1 = \unit{10}{\milli\second}$. Unsurprisingly, the performance of the chain is more affected for longer chains, although the reduction is much less drastic than for $T_2$ noise, and a finite amount of entanglement can be transferred even for long chains. 

\section{Coupling-strength disorder} 
In this section, we simulate disorder in the intra-chain coupling strength $\kappa$, as would arise from imprecisions when implanting the nitrogen impurity. For numerical convenience, we once more restrict our calculations to the single excitation subspace. We assume the spacings between the neighboring spins obey a Gaussian distribution around the mean value $r_{\text{N,N}} = \unit{10}{\nano\meter}$ ($\kappa = \unit{26}{\kilo\hertz}$). Our results are averages over a hundred independent runs for each data point. 

\begin{figure}[h]
\begin{center}
\includegraphics[width=2.65in]{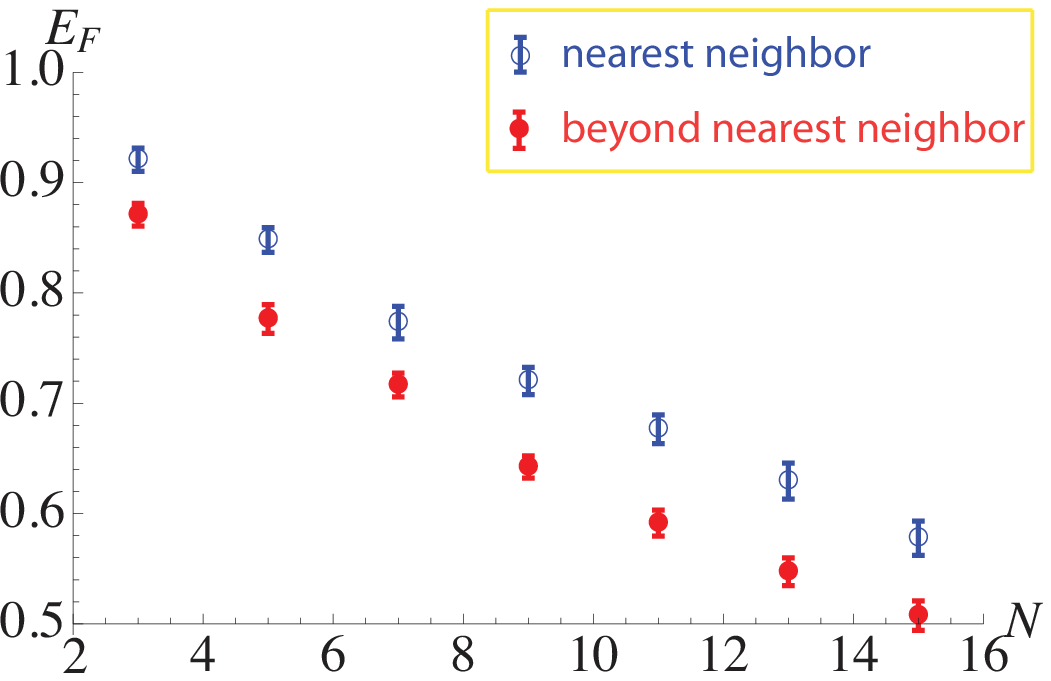}
\caption{Maximally achievable $E_F$ between the ancilla and the remote NV$^-$ qubit for different chain lengths $N$. The intra-chain spacings are assumed to follow a  Gaussian distribution with mean $r_{\text{N,N}} = \unit{10}{\nano\meter}$ with a standard deviation corresponding to $5\%$ disorder ($\sim$~15$\%$ disorder in the intra-chain coupling strength $\kappa$). The channel spins are initialised in the state $\ket{0}$, and each data point is the result of an average of 100 independent runs. The error bars indicate a $95\%$ confidence interval. The `nearest neighor' case uses the more practically relevant strong coupling regime $g=\kappa_{\text{mean}}$. For `beyond nearest neighbor' coupling, all pairwise spin couplings are included in accordance with the dipolar $1/r^3$ distance dependence.} 
\label{fig:kappadisorder}
\end{center}
\end{figure}

Fig.~\ref{fig:kappadisorder} shows that even a sizeable 15\% spread in the $\kappa$ distribution does not have a catastrophic impact on the entangling capacity of the chains. In fact, the reduction of achievable $E_F$ is rather similar to that obtained from a $T_1$ time around \unit{10}{\milli\second}. While the difference is not huge, the nearest-neighbor coupled chain proves consistently more robust.  

However, for the short chains considered in the main text, a small amount of disorder in $\kappa$ is unlikely to be the limiting factor preventing entanglement distribution. Ultimately, we expect the major challenge for this protocol is attaining sufficiently long $T_2$ times to overcome the limitations discussed in the main Letter.

\vspace{2mm}
\section*{Appendix: Rotating Wave Approximations}
\appendix
\setcounter{figure}{0}
\setcounter{equation}{0}
\renewcommand{\thefigure}{A\arabic{figure}}
\renewcommand{\theequation}{A\arabic{equation}}

Under the additional driving field of Eq.~(\ref{eq:driving}), the total Hamiltonian for two nitrogen electronic spin qubits reads
\begin{align}
&H^{\text{tot}}_{\text{N},\text{N}}  = \kappa\ \sz^1 \sz^2 + \sum_{i=1,2} \frac{\omega_i}{2} \sz^i + \sum_{i=1,2} \Omega\ \sx^i \cos \omega_i t \label{eq:HNNtot} \\ \nonumber
& =  \left( {\begin{array}{cccc}
\kappa + \frac{\omega_1 + \omega_2}{2} & \Omega \cos \omega_2 t & \Omega \cos \omega_1 t & 0 \\
\Omega \cos \omega_2 t & - \kappa + \frac{\omega_1 - \omega_2}{2} & 0 & \Omega \cos \omega_1 t \\
\Omega \cos \omega_1 t & 0 & - \kappa + \frac{\omega_2 - \omega_1}{2} & \Omega \cos \omega_2 t \\
0 & \Omega \cos \omega_1 t & \Omega \cos \omega_2 t & \kappa - \frac{\omega_1 + \omega_2}{2} \\
\end{array}}\right)~.
\end{align}
The unitary transformation for moving into the rotating frame is given by
\begin{equation}
U = \left( {\begin{array}{cccc}
e^{i \theta_1 t} & 0 & 0 & 0 \\
0 & e^{i \theta_2 t} & 0 & 0 \\
0 & 0 & e^{i \theta_3 t} & 0 \\
0 & 0 & 0 & e^{i \theta_4 t}  \\
\end{array}}\right)~,
\label{eq:unitary}
\end{equation}
where $\theta_{1 (4)} = (-) \frac{\omega_1 + \omega_2}{2}$ and $\theta_{2 (3)} = (-) \frac{\omega_1 - \omega_2}{2}$.  Applying the transformation $H_{\text{RF}}  =  i \dot{U} U^{\dagger} + U H U^{\dagger} $ and making the usual rotating wave approximation (RWA), justified since  $\Omega, |\kappa| \ll \omega_i$, yields:
\begin{align}
H_{\text{RWA}} &  = 
\left( {\begin{array}{cccc}
\kappa  & \frac{\Omega}{2} & \frac{\Omega}{2} & 0 \\
\frac{\Omega}{2} & - \kappa & 0 & \frac{\Omega}{2} \\
\frac{\Omega}{2} & 0 & - \kappa & \frac{\Omega}{2}\\
0 & \frac{\Omega}{2} & \frac{\Omega}{2} & \kappa  \\
\end{array}}\right) \nonumber \\ 
& = \kappa\ \sz^1 \sz^2 + \sum_{i=1,2} \frac{\Omega}{2} \sx^i~.
\label{eq:rf}
\end{align}

In the rotated basis with $(x, y, z) \rightarrow (z, -y, x)$ \big(i.e.~$\ket{0 (1)} \rightarrow \ket{+ (-)} =(\ket{0} \pm \ket{1})/\sqrt{2}$\big), the Hamiltonian~(\ref{eq:rf}) then becomes
\begin{align}
H_{\text{RF}} & =  \kappa\ \sx^1 \sx^2 + \sum_{i=1,2} \frac{\Omega}{2} \sz^i \nonumber \\
& =  \kappa\ (\sigma_+^1 + \sigma_-^1) (\sigma_+^2 + \sigma_-^2) + \sum_{i=1,2} \frac{\Omega}{2} \sz^i \nonumber \\
& \simeq  \kappa\ (\sigma_+^1 \sigma_-^2 + \sigma_-^1 \sigma_+^2) + \sum_{i=1,2} \frac{\Omega}{2} \sz^i~.
\label{eq:eff1}
\end{align}
Here, a second RWA was made in the last line by  neglecting the non-spin-conserving terms, which is valid when $|\kappa| \ll \Omega$. 

The same procedure can be generalised to a longer chain straightforwardly, allowing us to arrive at the desired nearest-neighbor interaction Hamiltonian~(\ref{eq:eff}).


\bibliographystyle{apsrev}